\documentclass{desyproc}

\usepackage{amsmath,amssymb,amsfonts}

\begin{document}

\title{\vspace{-3cm}
\hfill{\small\textnormal{IPPP/09/66; DCPT/09/132}}\\[2cm]
Minicharges, Monopoles, and Magnetic Mixing}

\author{{\slshape F.~Br\"ummer}\\[1ex]
Institute for Particle Physics Phenomenology, Durham University, DH1 3LE, UK
}

\contribID{bruemmer\_felix}

\desyproc{DESY-PROC-2009-05}
\acronym{Patras 2009} 
\doi  

\maketitle

\begin{abstract}
\noindent Minicharged particles (MCPs) arise naturally in extensions of the
Standard Model with hidden sector gauge groups. Many such extensions also
contain magnetic monopoles. For models containing both monopoles and MCPs, we
clarify the role of the Dirac charge quantization condition in restricting the
possible charges. We also show that monopoles of the hidden sector may manifest
themselves as MCPs, by a generalization of the Witten effect, which we call
\emph{magnetic mixing}.
\end{abstract}

\section{Introduction}

Many extensions of the Standard Model contain additional U(1) gauge factors as
part of a hidden sector. These may arise, for instance, directly from some
string compactification, or from a non-abelian gauge factor spontaneously broken
to U(1). For a single extra U(1), the most general low-energy Lagrangian for the
abelian gauge fields is then 
\begin{equation}\label{mostgeneralL}
\begin{split}
{\cal L}=&-\frac{1}{4}\left(F_{\mu\nu} F^{\mu\nu}+G_{\mu\nu}
G^{\mu\nu}+2\,\chi\,F_{\mu\nu} G^{\mu\nu}\right)\\
&-\frac{1}{32\pi^2}\left(\theta_F\,F_{\mu\nu} \tilde F^{\mu\nu}
+\theta_G\,G_{\mu\nu} \tilde G^{\mu\nu}+2\,\theta_{FG}\, F_{\mu\nu} \tilde
G^{\mu\nu}\right).
\end{split}
\end{equation}
Here $F_{\mu\nu}$ is the field strength of electromagnetism, $G_{\mu\nu}$ is the
field strength of the hidden sector U(1), $\tilde F_{\mu\nu}$ and $\tilde
G_{\mu\nu}$ are the respective dual field strengths, and $\chi$, $\theta_F$,
$\theta_G$, $\theta_{FG}$ are constants.

The first line represents the ordinary kinetic Lagrangian, including a kinetic
mixing term $2\,\chi\,F_{\mu\nu} G^{\mu\nu}$. If there are massive fields
charged under both $F_{\mu\nu}$ and $G_{\mu\nu}$, kinetic mixing is generically
induced radiatively \cite{Holdom:1985ag}. Fields which were charged under
$G_{\mu\nu}$ only will then pick up effective electromagnetic charges
$\sim\chi$, and will show up as minicharged particles (MCPs).\footnote{For some
concrete models in field theory and string theory, see
e.g.~\cite{Dienes:1996zr}.}

The $\theta$-terms in the second line are usually ignored in abelian theories.
They do not affect the equations of motion and carry no topological charges.
However, they do become important in the presence of magnetic monopoles: By the
Witten effect \cite{Witten:1979ey}, a $\theta_F\,F_{\mu\nu}\tilde F^{\mu\nu}$
term causes a monopole to pick up an electric charge $\sim\theta_F$. Similarly,
as we will argue, a \emph{magnetic mixing} term $\theta_{FG}\,F_{\mu\nu}\tilde
G^{\mu\nu}$ will cause a hidden sector monopole to pick up a visible electric
minicharge. 

In the following we will consider models with both kinetic and magnetic mixing
terms and with magnetic monopoles. We will show that the Dirac quantization
condition for electric charges must be suitably modified in the presence of
kinetic mixing, in order not to lead to a contradiction between charge
quantization and the appearance of MCPs \cite{Brummer:2009cs}. We will also
demonstrate how magnetic mixing terms may give electric minicharges to hidden
sector monopoles \cite{Bruemmer:2009ky}. Our considerations will in particular
apply to the case where the hidden sector U(1) is the remainder of a
spontaneously broken non-abelian gauge group, in which magnetic monopoles appear
naturally as 't Hooft--Polyakov monopoles
\cite{'tHooft:1974qc}.

\section{Kinetic mixing and charge quantization}

Let us start by ignoring the $\theta$-terms for now and consider a model with a
kinetic mixing term. For a U$(1)\times$U$(1)$ gauge theory with field strengths
$F=dA$ and $G=dB$, the Lagrangian is
\[
 {\cal L}=-\frac{1}{4}F_{\mu\nu} F^{\mu\nu}-\frac{1}{4}G_{\mu\nu}
G^{\mu\nu}-\frac{\chi}{2} F^{\mu\nu}G_{\mu\nu}-e j_\mu A^\mu -e' j'_\mu B^\mu.
\]
Diagonalizing the gauge kinetic terms, by defining  $C\equiv B+\chi A$ and
$H\equiv dC = G + \chi F$ and eliminating $B$ and $G$ in favour of $C$ and $H$,
gives a coupling of the current $j'$ to the gauge field $A$ with charge 
$-\chi e'$. Identifying $A$ with the photon of electromagnetism, hidden sector
charged matter fields have picked up ordinary electric charges $\sim\chi$, thus
becoming MCPs. If the MCPs are light, these induced charges must in fact be tiny
to evade experimental bounds \cite{Davidson:2000hf}.

In the presence of magnetic monopoles, electric charges should be quantized
\cite{Dirac:1931kp}. This can most easily be seen as follows: Consider a static
system consisting of an electron and a monopole. This system carries angular
momentum, which semi-classically should be quantized 
\cite{Wilson:1949}:
\[
  {\bf L}=\int d^3x\;{\bf x}\times\left({\bf E}\times{\bf
B}\right)=\frac{eg}{4\pi}\,{\bf n}.
\]
Here $e$ is the electron charge, $g$ is the monopole charge, and ${\bf n}$ is a
unit vector pointing from one towards the other. Requiring $|{\bf L}|$ to be
half-integral gives the Dirac quantization condition
\begin{equation}
eg\in 2\pi\mathbb{Z}.
\end{equation}
It follows that the ratio of any two electric charges $e_i$ and $e_j$ should be
rational, $e_{i}/e_{j}\in {\mathbb{Q}}$.

In models with a visible and a hidden sector U$(1)$, and with kinetic mixing
between the two, this leads to a problem because the induced minicharges are
proportional to $\chi$, which is an arbitrary and generally irrational number.
The problem is solved, however, if we only allow for monopoles which carry a
suitable magnetic charge also under the hidden U$(1)$. For example, consider a
model with an electron, an MCP, and a magnetic monopole with the following
electric and magnetic charges:
\begin{center}
\begin{tabular}
{|c|c|c|c|c|}\hline Particle & $q_{\rm vis}$ & $q_{\rm hid}$ & $g_{\rm vis}$ &
$g_{\rm hid}$\\\hline
electron & $e$ & $0$ & $0$ & $0$ \\
MCP & $-\chi e'$ & $e'$ & $0$ & $0$ \\
monopole & $0$ & $0$ & $g$ & $g'$\\ \hline
\end{tabular}
\end{center}
The total angular momentum of the combined hidden and ordinary electromagnetic
fields is, in a basis where the gauge-kinetic Lagrangian is diagonal,
\[
  {\bf L}=\int d^3x\;{\bf x}\times\left({\bf E}_{\rm vis}\times{\bf B}_{\rm
vis}+{\bf E}_{\rm hid}\times{\bf B}_{\rm hid}\right),\qquad |{\bf
L}|=\frac{q_{\rm vis}\,g}{4\pi}+\frac{q_{\rm hid}\,g'}{4\pi}.
\]
It is quantized if the monopole charges are $(g,g')=\left(0,\frac{2\pi
n}{e'}\right)$ or $(g,g')=\left(\frac{2\pi m}{e}, \frac{2\pi \chi m}{e}\right)$
or any linear combination of these, with $n,m\in\mathbb{Z}$. Only monopoles with
these quantum numbers can be consistently included in the model. This condition
on monopole charges is in fact a special case of the Schwinger--Zwanziger dyon
charge quantization condition \cite{Schwinger:1966nj} with multiple U$(1)$s.

In models with fundamental U$(1)$s, one may or may not choose to include
magnetic monopoles. By contrast, in many models where one of the U$(1)$s is the
remnant of a spontaneously broken non-abelian gauge group, magnetic monopoles
necessarily appear as topologically non-trivial field configurations.  As a
simple example consider a model where the hidden sector gauge group is SU$(2)$,
spontaneously broken to U$(1)$ by an adjoint scalar $\phi^a$ (the 't
Hooft--Polyakov model \cite{'tHooft:1974qc}). The Lagrangian for the scalar and
the hidden sector gauge field with field strength $G^a_{\mu\nu}$ is
\[
{\cal L}=-\frac{1}{4} G_{\mu\nu}^a G^{\mu\nu\,a} - \frac{1}{2}(D_\mu \phi)^a
(D^\mu\phi)^a+m^2\phi^a\phi^a-\lambda(\phi^a\phi^a)^2.
\]
A field configuration which represents a monopole at the origin, $r=0$, is given
by $\langle\phi^a\rangle=r^a\,f(r)$ with $f(r)$ a certain function. It breaks
SU$(2)\,\rightarrow$ U$(1)$ at large $r$. We can couple this model to a visible
sector U$(1)$ with field strength $F_{\mu\nu}$ by adding the terms
\[
\Delta {\cal L}=-\frac{1}{4}F_{\mu\nu}
F^{\mu\nu}-\frac{1}{2M}\phi^aG^{\mu\nu\,a}F_{\mu\nu}.
\]
The last term represents kinetic mixing between the surviving hidden U$(1)$ and
the visible U$(1)$. It is generated by integrating out heavy states which are
charged under both the visible and the hidden sector gauge group.
We have checked \cite{Brummer:2009cs} in this setting that the 't
Hooft--Polyakov monopole carries a monopole charge of precisely the allowed
kind, as by the above discussion.

\section{Magnetic mixing}

In a model with a single U$(1)$, a $\theta$-term $\sim F^{\mu\nu}\tilde
F_{\mu\nu}$ in the Lagrangian density
gives electric charges to magnetic monopoles \cite{Witten:1979ey}. This can be
seen as follows: Consider a magnetic monopole background with monopole charge
$g$ at $r=0$, superimposed with some static gauge potential $(A^\mu)=(A^0,{\bf
A})$. The $\theta$-term can be written in terms of electric and magnetic fields
as
\[
-\frac{\theta}{32\pi^2}F^{\mu\nu}\tilde F_{\mu\nu}=\frac{\theta}{8\pi^2}{\bf
E}\cdot{\bf B} = \frac{\theta}{8\pi^2}\left(\nabla
A^0\right)\cdot\left(\nabla\times{\bf A}+\frac{g}{4\pi}\frac{\bf
r}{r^3}\right).
\]
By integrating by parts, the Lagrangian contains a piece
\[
L=\int d^3r\,{\cal L}\supset -\frac{\theta}{8\pi^2}\int d^3r\,
A^0\,\nabla\cdot\frac{g\,\bf r}{4\pi\,r^3}=-\frac{\theta g}{8\pi^2}\int
d^3r\, A^0 \delta^3({\bf r}).
\]
This is a coupling of an electric point charge $-\theta g/(8\pi^2)$, located
at at ${\bf r}=0$, to the electrostatic potential $A^0$. In other words, the
monopole has acquired an electric charge.

In a model with a visible U$(1)$ and a hidden U$(1)$, and a magnetic mixing term
\[
{\cal L}\supset-\frac{\theta_{FG}}{32\pi^2}F_{\mu\nu}\tilde G^{\mu\nu}
\]
such as in Eq.~\eqref{mostgeneralL}, an analogous calculation
\cite{Bruemmer:2009ky} shows that hidden magnetic monopoles acquire visible
electric charges. This is potentially very interesting with regard to
phenomenology: Magnetic monopoles of ordinary electromagnetism are expected to
be much heavier than $M_{\rm GUT}$ and therefore undetectable. Hidden sector
monopoles, on the other hand, could very well be relatively light. If they carry
electromagnetic charges from magnetic mixing, they could be detected in the same
way as MCPs. 

For instance, if the monopole is again a 't Hooft--Polyakov monopole of a
spontaneously broken non-abelian gauge group in the hidden sector, its mass is
semi-classically given by the breaking scale, divided by the gauge coupling. For
models with an arbitrarily low breaking scale, the monopole could be arbitrarily
light. Alternatively, if a high breaking scale is preferred for naturalness
reasons, one might speculate that the hidden sector gauge group could be
strongly coupled, such that the semi-classical approximation is invalid and
monopoles could still be light. In fact, with the Seiberg--Witten model
\cite{Seiberg:1994rs} there exists even a calculable example of a gauge theory 
with 't Hooft--Polyakov monopoles becoming arbitrarily light in some 
strong-coupling region of moduli space.

Despite the fact that there is no non-abelian analogue of kinetic or magnetic
mixing terms, they may be generated radiatively \cite{Bruemmer:2009ky} once a 
non-abelian gauge group is broken to U$(1)$ (which is precisely the situation in 
which 't Hooft--Polyakov monopoles appear). For inducing these terms, the model 
should contain matter fields charged under both the visible and the hidden 
sector. Magnetic mixing can only be induced if these fields possess CP-violating
couplings, since a magnetic mixing term itself violates CP.

\section*{Acknowledgments}
The author would like to thank J.~Jaeckel and V.~V.~Khoze for collaboration in
\cite{Brummer:2009cs,Bruemmer:2009ky}, and the organizers of the \emph{5th
Patras Workshop} for a pleasant and interesting meeting.

\begin{footnotesize}

\end{footnotesize}

\end{document}